\documentclass[nofootinbib,prd,reprint,preprintnumbers,twocolumn]{revtex4}

\usepackage{amsmath,amssymb,bm,mathrsfs}
\usepackage{graphicx,color}
\usepackage{amsfonts}
\usepackage[export]{adjustbox}
\usepackage{mathtools}
\usepackage[sort&compress]{natbib} 
\usepackage{srcltx}
\usepackage{dsfont}
\usepackage{epsfig}
\usepackage{bbold}
\usepackage{rotating}
\usepackage[dvipsnames]{xcolor}
\usepackage{subfig}
\usepackage{xfrac}
\usepackage{multirow}
\usepackage{booktabs}
\usepackage{ulem}
\usepackage{siunitx}
\usepackage{mwe}
\usepackage[inline]{enumitem} % for inline list
\usepackage[colorlinks=true
,urlcolor=blue
,anchorcolor=blue
,citecolor=blue
,filecolor=blue
,linkcolor=red
,menucolor=blue
,linktocpage=true
,pdfproducer=medialab
,pdfa=true
]{hyperref}
\usepackage{cleveref}
\usepackage{setspace}

\usepackage{caption}
\usepackage{tikz}

\definecolor{forestgreen}{RGB}{34,139,34}

\newcounter{qnumber}

\begin{document}

%=============================================================================

\title{Rogue worlds meet the dark side: revealing terrestrial-mass primordial black holes with the Nancy Grace Roman Space Telescope}

\author{William DeRocco}

\author{Evan Frangipane}

\author{Nick Hamer}

\author{Stefano Profumo}

\author{Nolan Smyth}

\affiliation{Department of Physics, University of California Santa Cruz, 1156 High St., Santa Cruz, CA 95064, USA\\
and Santa Cruz Institute for Particle Physics, 1156 High St., Santa Cruz, CA 95064, USA}
%=============================================================================

%-----------------------------------------------------------------------------
\begin{abstract}
Gravitational microlensing is one of the strongest observational techniques to observe non-luminous astrophysical bodies. Existing microlensing observations provide tantalizing evidence of a population of low-mass objects whose origin is unknown. These events may be caused by terrestrial-mass free-floating planets or by exotic objects such as primordial black holes. However, the nature of these objects cannot be resolved on an event-by-event basis, as the induced light curve is degenerate for lensing bodies of identical mass. One must instead statistically compare \textit{distributions} of lensing events to determine the nature of the lensing population. While existing surveys lack the statistics required to identify multiple subpopulations of lenses, this will change with the launch of the Nancy Grace Roman Space Telescope. Roman's Galactic Bulge Time Domain Survey is expected to observe hundreds of low-mass microlensing events, enabling a robust statistical characterization of this population. In this paper, we show that by exploiting features in the distribution of lensing event durations, Roman will be sensitive to a subpopulation of primordial black holes hidden amongst a background of free-floating planets. Roman's reach will extend to primordial black hole dark matter fractions as low as $f_\text{PBH} = 10^{-4}$ at peak sensitivity, and will be able to conclusively determine the origin of existing ultrashort-timescale microlensing events. A positive detection would  provide evidence that a significant fraction of the cosmological dark matter consists of macroscopic, non-luminous objects.
\end{abstract}

\maketitle
\section{Introduction}
\label{sec:intro}

The nature of dark matter remains one of the most pressing open questions in fundamental physics. While multiple lines of compelling evidence indicate its existence, its microphysical nature remains unknown (for a recent review and up to date references see e.g. Chapter 27 of Ref.~\cite{ParticleDataGroup:2022pth}). Many models have been proposed to explain this additional matter content, with many such models  introducing new fundamental particles with suppressed interaction cross-sections to populate the dark sector \cite{ParticleDataGroup:2022pth}. However, dark matter may instead be macroscopic and potentially possess large interaction cross-sections, escaping detection due to its low number density. Primordial black holes (PBHs) are a well-motivated candidate for such a macroscopic dark matter model \cite{Hawking:1974rv, Carr:1974nx, Carr:2020gox, Carr:2021bzv}. There are a wide variety of mechanisms that result in the formation of PBHs, from the collapse of overdensities sourced by inflation \cite{Hawking:1971ei,Carr:1974nx} to phase transitions \cite{Khlopov:1980mg} and topological defect collapse \cite{Crawford:1982yz} in the early universe (see the discussion in Sec.~\ref{sec:pbhs} below). PBHs may form over a wide range of masses, from as low as asteroid masses up to thousands of solar masses and beyond.

The Earth-mass range, $\sim 10^{-6}\,M_\odot$, is of particular interest, as observations of excess short-duration microlensing events have been suggested to constitute a first hint of a population of PBHs at terrestrial masses \cite{Niikura_2019}.
However, there is another possible candidate to explain these events: free-floating planets (FFPs). These are planets that have been ejected from their parent star system by dynamical interactions during the chaotic early phases of system formation. Such FFPs are expected to dramatically outnumber bound exoplanets at sub-terrestrial masses \cite{Strigari_2012,sumi2023freefloating}, constituting a large potential background for surveys seeking to observe PBHs at Earth masses and below.

Previously, constraints on the PBH abundance have been placed in regions of parameter space for which the expected contribution from FFPs is negligible. 
However, with the upcoming launch of the Nancy Grace Roman Space Telescope, this will change: over the course of its Galactic Bulge Time Domain Survey (GBTDS) \cite{2022BAAS...54e.146G}, Roman is expected to observe hundreds of free-floating planets at roughly Mars mass and above \cite{Johnson_2020}. This unprecedented sensitivity will also provide the opportunity to search for PBHs in new regions of parameter space. In these regions, FFPs constitute an irreducible background that must be taken into account in order to constrain or claim the discovery of PBHs. 

FFPs and PBHs cannot generally be discriminated on an event-by-event basis, as their light curves are degenerate for identical masses. However, FFPs and PBHs are expected to arise from different underlying mass distributions, permitting a statistical means of discrimination. In this paper, we present a method by which a subpopulation of PBHs can be detected amidst a background of FFPs. We find that even in the presence of FFPs, Roman will be sensitive to PBHs at abundances well below existing constraints. In particular, Roman will be able to conclusively determine the nature of the Earth-mass ``hint'' of a PBH population claimed by \cite{Niikura_2019}.

The remainder of the paper is organized as follows. In Sec. \ref{sec:microlensing}, we discuss microlensing surveys and describe the observables associated with microlensing lightcurves. In Sec.~\ref{sec:targets}, we review mechanisms for PBH/FFP formation and provide a fiducial mass function for the abundance of each population. In Sec. \ref{sec:methods}, we describe the implementation of our analysis framework and the associated statistical methodology for estimating Roman sensitivity. In Sec. \ref{sec:disc}, we present our results and discuss their implications before concluding in Sec. \ref{sec:conc}.

\section{Microlensing}
\label{sec:microlensing}

Gravitational lensing is a powerful technique to observe non-luminous massive objects at astronomical distances. Light rays passing in the vicinity of a massive object are bent by the gravitational field of the object, causing the light from background stars (``sources'') to be distorted by massive objects (``lenses'') that lie along the line of sight. For high mass lenses, this effect produces multiple images of the source; for low mass lenses, the images cannot be individually resolved and instead contribute to an overall apparent magnification of the source. This effect is known as microlensing \cite{paczynski_gravitational_1986}. 

The duration and magnification of the source are determined by the mass of lens $M$, the distance to the lens and source, $d_L$ and $d_S$, the relative proper motion of the source and the lens $\mu_\text{rel}$, the impact parameter $u$, the angular diameter of the source $\theta_S$, and the effective angular diameter of the lens $\theta_E$. This final quantity is also known as the ``Einstein angle'' and is given by 
\begin{equation}
\label{eq:thetaE}
\theta_E = \sqrt{\frac{4 G M (1 - d_L/d_S)}{c^2 d_L}}.
\end{equation}

When $\theta_S \ll \theta_E$, the angular extent of the source is negligible. This ``point-source regime'' is typical for large lens masses and distant sources, and the associated event duration is given by the time it takes for the source to cross the Einstein radius of the lens. This ``Einstein crossing time'' is defined as
\begin{equation}
\label{eq:tE}
    t_E = \frac{\theta_E}{\mu_\text{rel}}. 
\end{equation}
In the point-source regime, the apparent magnification is given by
\citep{nakamura_wave_1999} 
\begin{equation}
    A_\text{ps}(u) = \frac{u^2 + 2}{u\sqrt{u^2 + 4}},
    \label{Ageo}
\end{equation}
where $u \equiv u(t)$ is the impact parameter as a function of time. This yields a characteristic light curve consisting of a narrow peak.

When $\theta_S \gtrsim \theta_E$, however, the point-source approximation breaks down. In this finite-source regime, the light curve saturates at a lower maximum magnification and the event duration is no longer set by $t_E$, but rather by the time for the \textit{lens} to cross the finite angular extent of the \textit{source}, a timescale of $\sim 2\theta_S/\mu_\text{rel}$. Similarly, the magnification in this regime no longer diverges as $u\rightarrow 0$ and is instead given by an integral over the source disk, specified in polar coordinates $(r,\phi)$ as
\citep{matsunaga_finite_2006, witt_can_1994, sugiyama_revisiting_2019}
\begin{multline}
    A_{\mathrm{finite}}(u,\rho) \equiv \\
    \frac{1}{\pi \rho^2} \int_0^{\rho} dr \int_0^{2\pi} d\phi ~r ~A_{\rm{ps}}\Big( \sqrt{r^2 + u^2 - 2ur \cos(\phi)} \Big),
    \label{Afinite}
\end{multline}
where $\rho \equiv \theta_S/\theta_E$ and the origin has been chosen such that the lens center is located at a distance $u$ from the origin along $\phi=0$. The maximum impact parameter that produces a detectable event is defined implicitly via the relation $A_{\mathrm{finite}}(u_{\rm{T}},\rho) = A_T$, where the minimum detectable magnification, $A_T$, is set by the photometric sensitivity of the microlensing survey, and $u_T$ is the maximal impact parameter that results in a magnification of at least $A_T$. $u_T$ defines the phase space for the expected event rate calculation (see Sec. \ref{subsec:eventrate}) and can be calculated for a given $d_L, d_S$, and $\theta_S$ following the procedure given in \cite{sugiyama_revisiting_2019}.

For most events, the fundamental observable that can be measured from the light curve is the duration. We define this as the time over which the magnification is above detection threshold ($A > A_T$ or equivalently $u < u_T$): 
\begin{equation}
\label{eq:tdur}
    t_\text{dur} = 2 \sqrt{u_T^2 - u_{\rm{min}}^2}  \frac{\theta_E}{\mu_\text{rel}},
\end{equation} 
where $u_\text{min}$ is the impact parameter at the point of closest approach. Assuming perfect photometry, $u_T \approx \rho$ in the extreme finite-source regime and $\approx 1$ in the point-source regime; hence, for a trajectory that passes through the midplane of the source,  $t_\text{dur}$ approaches the expected $\sim 2\theta_S/\mu_\text{rel}$ in the finite-source limit and $\sim 2t_E$ in the point-source regime.

Though finite-source effects reduce the peak magnification, which can reduce detectability, they introduce characteristic features in the light-curve that permit a measurement of $\theta_E$. Coupled with a measurement of the lens distance, an estimate of the lens mass can be made. However, the extraction of $\theta_E$ is a challenge for many events, especially those that do not conform to simple single-lens models. 
Additionally, for low masses and short event durations, estimating $d_L$ requires a simultaneous observation by another telescope in order to provide a parallax measurement, which is often unavailable. As such, the only observable quantity that can be robustly measured for most microlensing events is the event duration, $t_\text{dur}$, and is therefore the quantity we choose to employ to discriminate amongst various subpopulations of lenses in Sec.~\ref{sec:methods}.

\section{Microlensing Targets}
\label{sec:targets}

In this section, we discuss two primary targets for microlensing surveys in the terrestrial mass range and connect them to existing observations.

\subsection{Primordial black holes}
\label{sec:pbhs}

Black holes not originating from the collapse of massive stars are generically termed ``primordial'' black holes  and appear in many extensions of the Standard Model. Most formation mechanisms rely upon the growth of large density fluctuations in the early universe that ultimately collapse. These may be seeded by features in the inflationary potential \cite{Carr:1993aq,Ivanov:1994pa,Yokoyama:1998pt,Garcia-Bellido:2017mdw,Ballesteros:2017fsr,Hertzberg:2017dkh,Germani:2017bcs} or by other physical processes, such as the collapse of inhomogeneities during the matter-dominated era triggered by a sudden pressure reduction \cite{Khlopov:1980mg, Jedamzik:1996mr, Byrnes:2018clq}, collapse of cosmic string loops \cite{Hawking:1987bn, Helfer:2018qgv, James-Turner:2019ssu}, bubble collisions \cite{Crawford:1982yz, Kusenko:2020pcg}, a scalar condensate collapsing to Q-balls before decay \cite{Cotner:2017tir, Cotner:2016cvr, Cotner:2018vug, Cotner:2019ykd}, or domain walls \cite{Rubin:2001yw, Ge:2019ihf, Deng:2017uwc, Liu:2019lul}. (See, e.g., \cite{Carr:2020gox,Carr:2021bzv} for recent reviews.)

If the overdensities are seeded by inflationary features, the resulting PBH masses are related to the redshift of formation since PBHs acquire a mass of order the total energy within a Hubble volume at the time of collapse. The resulting mass distribution is often well-described by a log-normal distribution, which is a generic prediction for PBHs forming from smooth, symmetric peaks in the power spectrum of density fluctuations in the early universe \cite{Dolgov:1992pu}. Numerical and analytical evidence for this functional form was provided in \cite{Green:2016xgy} and \cite{Kannike:2017bxn}, see also the recent Ref. \cite{Kleban:2023ugf}.
For this reason, in the following, we will consider a fiducial PBH mass function of the form 
\begin{equation}
\label{eq:pbh_mass_func}
    f(M,\sigma,M_c)=\frac{f_{\rm PBH}}{\left(\sqrt{2\pi}\right) \sigma M}\exp\left(-\frac{\log^2\left(M/M_c\right)}{2\sigma^2}\right),
\end{equation}
normalized such that 
\begin{equation}
    f_{\rm PBH}=\frac{\Omega_{\rm PBH}}{\Omega_{\rm DM}}=\int\ {\rm d}M f(M,\sigma,M_c)
\end{equation}
where here $\Omega_\text{PBH}$ and $\Omega_\text{DM}$ are the fractional energy density of PBHs and of all dark matter, respectively.
Here $M_c$ is the mean value of $M$ and $\sigma$ is the standard deviation of the logarithmic mass.

PBHs are a compelling candidate for dark matter and have been searched for across a wide range of masses. In the mass range of $\approx 10^{-11}\,M_\odot - 10\,M_\odot$, gravitational lensing sets some of the strongest observational constraints on their abundance \cite{Niikura_2019, Karami:2016rjp, Griest:2013esa, CalchiNovati:2013jpj, Wyrzykowski:2010mh, Wyrzykowski:2011tr, MACHO:2000bzs, MACHO:2000qbb} limiting the fractional energy density to $f_\text{PBH} \approx 10^{-1} - 10^{-2}$. At terrestrial masses, the strongest limits are set by observations made by the Optical Gravitational Lensing Experiment (OGLE) \cite{Niikura_2019}. However, this survey also revealed an anomalous excess of six short-duration events consistent with a population of Earth-mass PBHs at $f\approx 10^{-2}$. To date, the nature of these observations has not been resolved. As we will show in Sec.~\ref{sec:disc}, upcoming observations by the Nancy Grace Roman Space Telescope will be able to establish whether a population of PBHs truly exists at these masses or whether these events were more likely caused by, e.g., free-floating planets.

\subsection{Free-floating planets}
\label{sec:ffps}

The term ``free-floating planets'' is often used to describe two different classes of astrophysical objects. At masses near and above that of Jupiter, FFPs may form \textit{in situ} as the core of a failed star \cite{Miret_Roig_2023}. At lower masses, FFPs are expected to primarily form within young planetary systems before being ejected by dynamical processes onto unbound orbits. There is a wide variety of processes that can result in the ejection of a protoplanetary object, including stripping by nearby stars,  gravitational scattering off of planetesimals, and interactions with an inner binary star system \cite{fitzsimmons2023interstellar,doi:10.1126/science.274.5289.954,10.1111/j.1365-2966.2010.17730.x}. Both simulations and observations suggest that FFPs may dramatically outnumber bound planets at masses $\lesssim M_\oplus$ \cite{Strigari_2012,sumi2023freefloating, mroz2023exoplanet}. FFPs are therefore an exciting observational target for existing and upcoming microlensing surveys.

Ejection processes typically yield a distribution of FFPs that are well-described by a power law \cite{mroz2023exoplanet}. Here we adopt the form 
\begin{equation}
\label{eq:ffp_mass_func}
\frac{dN}{d \log_{10} (M)} = \mathcal{N} \Big( \frac{M}{M_{\rm{norm}}}\Big)^{-p}
\end{equation}
where $\mathcal{N}$ is the total number of FFPs per star at mass $M$ scaled by a  normalization mass $M_{\text{norm}}$. Throughout the rest of the paper, we take all logarithms to be base 10 and  $M_{\rm{norm}} = M_{\oplus}$ unless otherwise noted. 

At present, observational measurements of the FFP population do not place strong constraints on the values of $\mathcal{N}$ and $p$. Existing microlensing surveys have observed tens of FFPs, with only three events permitting a mass estimate placing the lens in the terrestrial range.\footnote{The associated events are OGLE-2012-BLG-1323 \citep{2019A&A...622A.201M}, OGLE-2016-BLG-1928 \citep{Mroz_2020}, and MOA-9y-5919 \citep{koshimoto2023terrestrial}.} Based off these data, combined with the results from simulations of ejection \citep{1968IAUS...33..486D,2009ApJ...704..733M,G_sp_r_2012} and observations of bound systems \citep{Strigari_2012,2019arXiv190603270S,Landgraf_2000}, the best estimates for $p$ and $\mathcal{N}$ are of order $p\approx 1$ and $\mathcal{N} \approx 10$ with an uncertainty spanning $p \approx 0.66 - 1.33$ and $\mathcal{N} \approx 2 - 20$ \cite{sumi2023freefloating,10.5303/JKAS.2022.55.5.173,mroz2023exoplanet}. We choose to adopt $p = 1$ and $\mathcal{N} = 10$ as our fiducial parameters and marginalize over the uncertainty on their values  when computing our sensitivity (see Sec.~\ref{subsec:AD}).

\section{Detecting PBHs with Roman}
\label{sec:methods}

In this section, we describe our statistical methodology for detecting a subpopulation of PBH lenses within a background of FFPs. 
The key point is that though PBH and FFP events cannot be discriminated on an event-by-event basis, the two \textit{populations} can be distinguished by the statistical distribution of their event durations, $t_\text{dur}$ (Eq. \ref{eq:tdur}). This distribution is predominantly controlled by the underlying mass function of the lensing population, which differs significantly between FFPs and PBHs (see Secs. \ref{sec:pbhs} and \ref{sec:ffps}). Additionally, the $t_\text{dur}$ distribution is influenced by the distribution of lens distances and transverse velocities, both of which differ between FFPs and PBHs as well (see Sec. \ref{subsec:eventrate}). As a result, the observed distribution of $t_\text{dur}$ provides a robust means of identifying multiple populations of lenses within a set of microlensing events.\footnote{An alternate strategy, as suggested by Niikura et al. \cite{Niikura_2019}, would be to observe along different lines of sight, as FFPs and PBHs are expected to follow different spatial distributions. As this would require an additional dedicated survey, we leave the study of this topic to future work.}

While existing observations have not yet yielded a sufficient number of detections at terrestrial masses to resolve the underlying distribution of $t_\text{dur}$, this will change in the coming years. The Galactic Bulge Time Domain Survey (GBTDS), one of three primary surveys to be conducted by the upcoming Nancy Grace Roman Space Telescope (set to launch in 2027), will observe seven fields tiling 2 square degrees of the Galactic bulge with a cadence of 15 minutes during six 72-day observing seasons \cite{2022BAAS...54e.146G}. This survey strategy has been designed specifically to meet core science requirements for the mission, including measuring the abundance of free-floating planets to within 25\%. As such, the GBTDS is expected to yield hundreds of FFP microlensing events at Mars mass and above \cite{Johnson_2020}, providing the opportunity for distribution-level analyses.

In the following two subsections, we will describe our methodology for determining Roman's sensitivity to discriminating a PBH subpopulation from a background FFP population using the observed distribution of $t_\text{dur}$ values. This is done in two steps. First, in Sec. \ref{subsec:eventrate}, we compute the event rate for both of these populations given Roman's fiducial survey parameters to determine the $t_\text{dur}$ distribution for both populations. Then, in Sec. \ref{subsec:AD}, we perform a 2-Sample Anderson-Darling test to determine the statistical significance at which a combined FFP and PBH $t_\text{dur}$ distribution differs from a FFP distribution without PBHs.

\subsection{Event rate estimation}
\label{subsec:eventrate}

The key input to our statistical methodology is the distribution of event durations, $t_\text{dur}$. In order to compute this, we integrate over the differential event rate given by \cite{batista_moa-2009-blg-387lb_2011, niikura_microlensing_2019}
\begin{multline}
    \label{eq:difRateAll}
     \frac{d\Gamma}{dM\,dd_L\,dt_\text{dur}\,du_\text{min}} =  \\
     \frac{2}{\sqrt{u_T^2 - u_{\rm{min}}^2}} \frac{v_T^4}{v_c^2} \exp \Big[ -\frac{v_T^2}{v_c^2}\Big] \frac{\rho_M}{M}f(M) \varepsilon(t_\text{dur}),
\end{multline}
where $f(M)$ is the probability distribution of lens masses (Eq. \ref{eq:pbh_mass_func} or Eq. \ref{eq:ffp_mass_func} for PBHs and FFPs, respectively), 
$\rho_M$ is the mass density of the lens population, 
$\varepsilon(t_\text{dur})$ is the detection efficiency, and $v_T$, the transverse velocity, is given by
\begin{equation}
\label{eq:vT}
 v_T = 2\theta_E d_L \sqrt{u_T^2 - u_{\min}^2}/t_\text{dur}.
\end{equation}
We set $u_T$, the maximum impact parameter that produces a detectable event, according to the procedure discussed in Sec.~\ref{sec:microlensing}, adopting $A_T = 1.34$ as our fiducial threshold magnification. This choice likely underestimates Roman's sensitivity, but is in keeping with the literature \cite{Johnson_2020} (see also App. \ref{sec:yields}). 
The event rate, $\Gamma$, is then evaluated as

\begin{multline}
    \label{eq:rate}
     \Gamma= 2 \int_{M_{\rm{min}}}^{M_{\rm{max}}} dM \int_0^{d_s}dd_L \int_0^{u_T}du_{\rm{min}} \int_{t_\text{min}}^{t_\text{max}}dt_\text{dur}\\
     \frac{1}{\sqrt{u_T^2 - u_{\rm{min}}^2}} \frac{v_T^4}{v_c^2} \exp \Big[ -\frac{v_T^2}{v_c^2}\Big] \frac{\rho_M}{M}f(M) \varepsilon(t_\text{dur}),
\end{multline}
which we calculate using \texttt{LensCalcPy},\footnote{https://github.com/NolanSmyth/LensCalcPy} a package to semi-analytically calculate microlensing observables. We take $t_\text{min}$ to be $15 ~\rm{min}$ and $t_\text{max}$ to be $6\times 72 ~\rm{days}$, corresponding to the proposed cadence and observational duration of the Roman GBTDS. By performing the integral and multiplying the resulting rate by the GBTDS observational duration, we compute the expected total number of events that Roman will detect, denoted $N_\text{FFP}$ and $N_\text{PBH}$ for FFPs and PBHs, respectively.

In computing the event rate, we must specify the velocity and spatial distributions of the lenses. We assume that the FFP density tracks the stellar distribution of the galaxy, for which we adopt the exponential Koshimoto parametric model described in \cite{koshimoto_parametric_2021}. We take the PBH mass distribution to be a Navarro-Frenk-White profile given by
\begin{equation}
    \label{eq:nfw}
    \rho_M = \frac{\rho_0}{(\frac{r}{r_s})(1 + \frac{r}{r_s})^2},
\end{equation}
where $\rho_0 = 4.88 \times 10^6 ~M_{\odot} ~\rm{kpc}^{-3}$ and $r_s = 21.5 ~\rm{kpc}$ \cite{klypin_lcdm-based_2001}.  While the relative source-lens velocity depends in general on the positions of both source and lens, we take $ v_c = 220 ~\rm{km/s}$ for PBHs and $v_c = 200 ~\rm{km/s}$ for FFPs. The former is a typical value for a virialized DM halo \cite{klypin_lcdm-based_2001}, and the latter is approximately the average transverse velocity in the stellar disk (see e.g. \cite{Niikura_2019} for a more complete description). As the majority of sources are in the Galactic Bulge, finite-source effects imply that the low-mass lenses we consider must be sufficiently far from the source in order to be detectable, making this simplification appropriate for the scope of this work. Ultimately, our results are fairly insensitive to changes in these choices of parameters, as the dominant uncertainty in our analysis arises from the normalization of the FFP mass function (see Sec.~\ref{sec:disc}). However, we have compared our yields to those computed by \cite{Johnson_2020}, which employ a different Galactic model and mass function, and find $\mathcal{O}(1)$ agreement (see App.~\ref{sec:yields}).

For the mass function of PBHs, we assume a log-normal distribution (Eq. \ref{eq:pbh_mass_func}), while for FFPs, we adopt a power-law (Eq. \ref{eq:ffp_mass_func}) truncated at $M_{\rm{min}} = 10^{-13}\,M_\odot$ and $M_{\rm{max}} = 0.1\,M_\odot$ for computational purposes. These cutoffs have been chosen to lie well outside the mass range of Roman's sensitivity ($\approx 10^{-8}\, M_\odot - 10^{-3}\, M_\odot$) and we have verified numerically that they do not have an effect on the results. 

\begin{figure}[b]
\includegraphics[width=\linewidth]{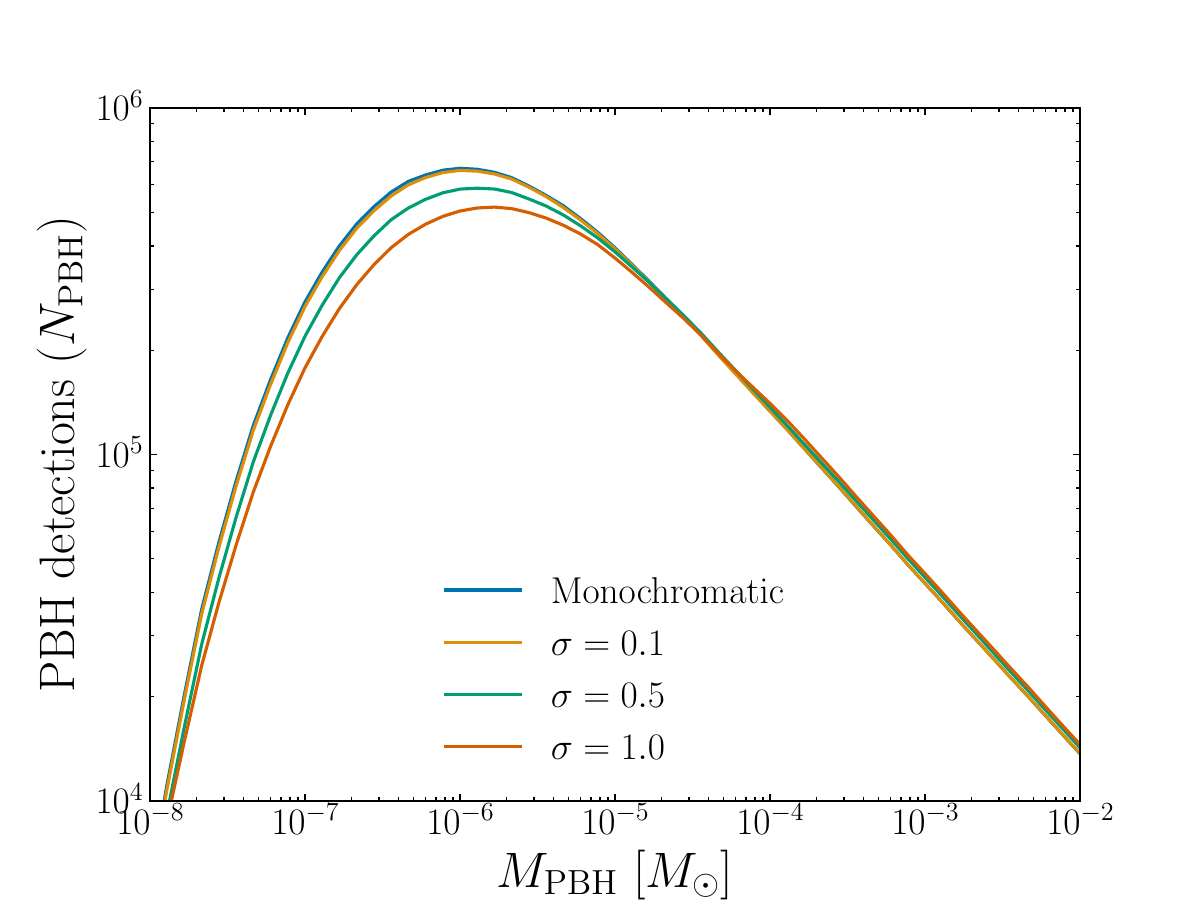}
\caption{The total number of PBH microlensing events detectable by Roman for $f_\text{PBH} = 1$ as a function of $M_\text{PBH}$. The different curves correspond to different widths of the PBH mass distribution (see Sec. \ref{sec:pbhs}).}
\label{fig:pbh_yield} 
\end{figure}

\begin{figure}[b]
\includegraphics[width=\linewidth]{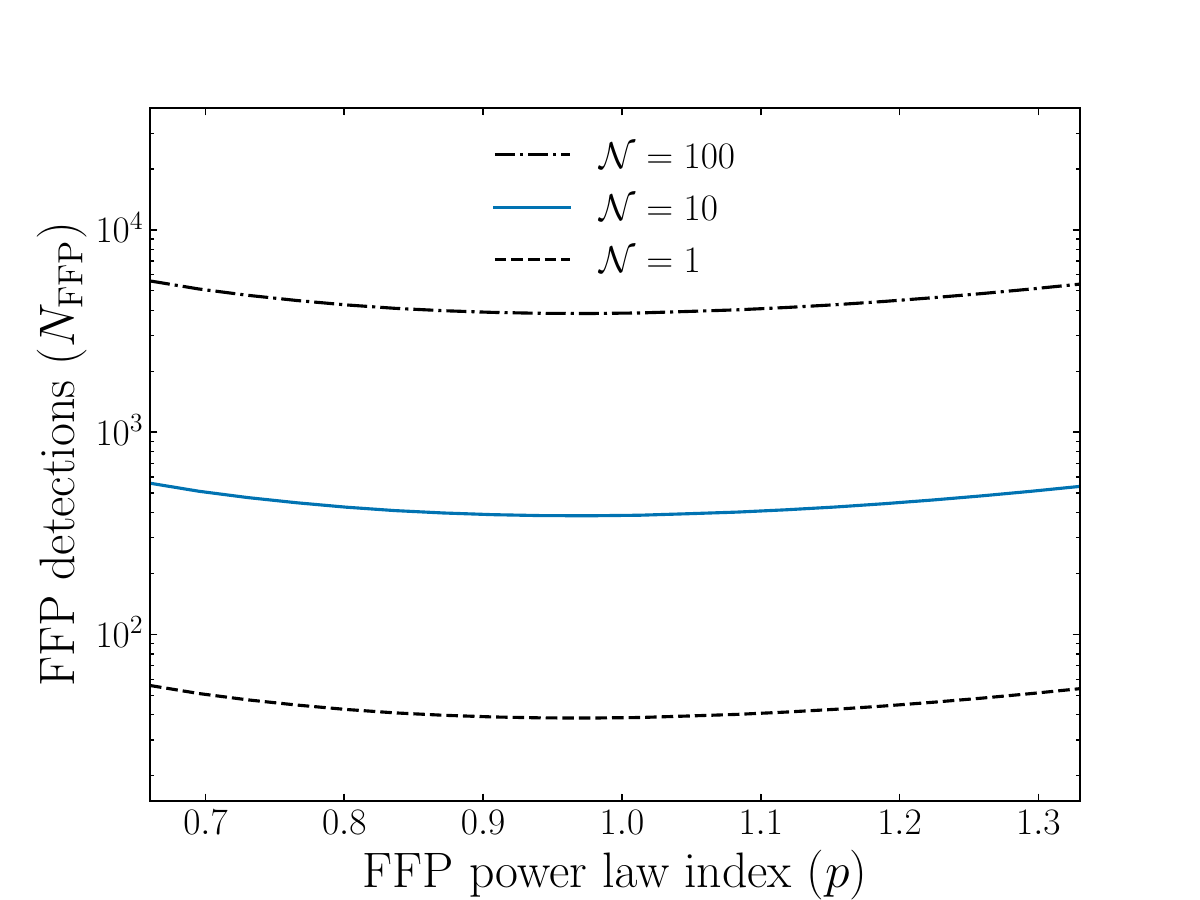}
\caption{The total number of FFP microlensing events detectable by Roman  as a function of $p$. The fiducial normalization $\mathcal{N} = 10$ is shown as a solid blue line, with $\mathcal{N}=1$ and 100 shown as dashed and dash-dotted curves, respectively.}
\label{fig:ffp_yield} 
\end{figure}

The resulting yields for PBHs and FFPs are shown in Figs. \ref{fig:pbh_yield} and \ref{fig:ffp_yield}. Fig. \ref{fig:pbh_yield} shows the number of PBH events Roman is expected to see during its proposed observational duration as a function of $M_\text{PBH}$ for $f_\text{PBH} = 1$. The various curves correspond to different widths of the log-normal distribution, $\sigma$.  Note that a $f_\text{PBH} = 1$ abundance has already been ruled out by other microlensing surveys, hence the yields in unconstrained parameter space are necessarily smaller than the values in Fig.~\ref{fig:pbh_yield}. We see that in unconstrained parameter space ($f \lesssim 10^{-2}$), Roman is expected to observe up to $\approx 10^4$ PBH events.\footnote{We note that though distinguishing FFPs from PBHs requires a statistical characterization when the observed yields of each are comparable, there are regions of parameter space in which PBH observations would well exceed the expected FFP yield, hence an identification of this population would be much simpler. Interestingly, this includes the parameter space in which PBHs explain the short-duration OGLE events, making their interpretation as FFPs more challenging.}

Fig. \ref{fig:ffp_yield} shows the number of FFP events Roman is expected to see during its proposed observational duration as a function of $p$, the power-law index of the FFP mass distribution. The various curves correspond to various normalizations of the power law, with $\mathcal{N} = 10$ the fiducial value. The yield is only weakly-dependent on $p$, with our fiducial distribution yielding $\approx 400$ events for a broad range of $p$.

\begin{figure*}
\hspace*{\fill}
\centering
\begin{minipage}[t]{0.48\textwidth}
\centering
\vspace{0pt}
\includegraphics[width=\linewidth]{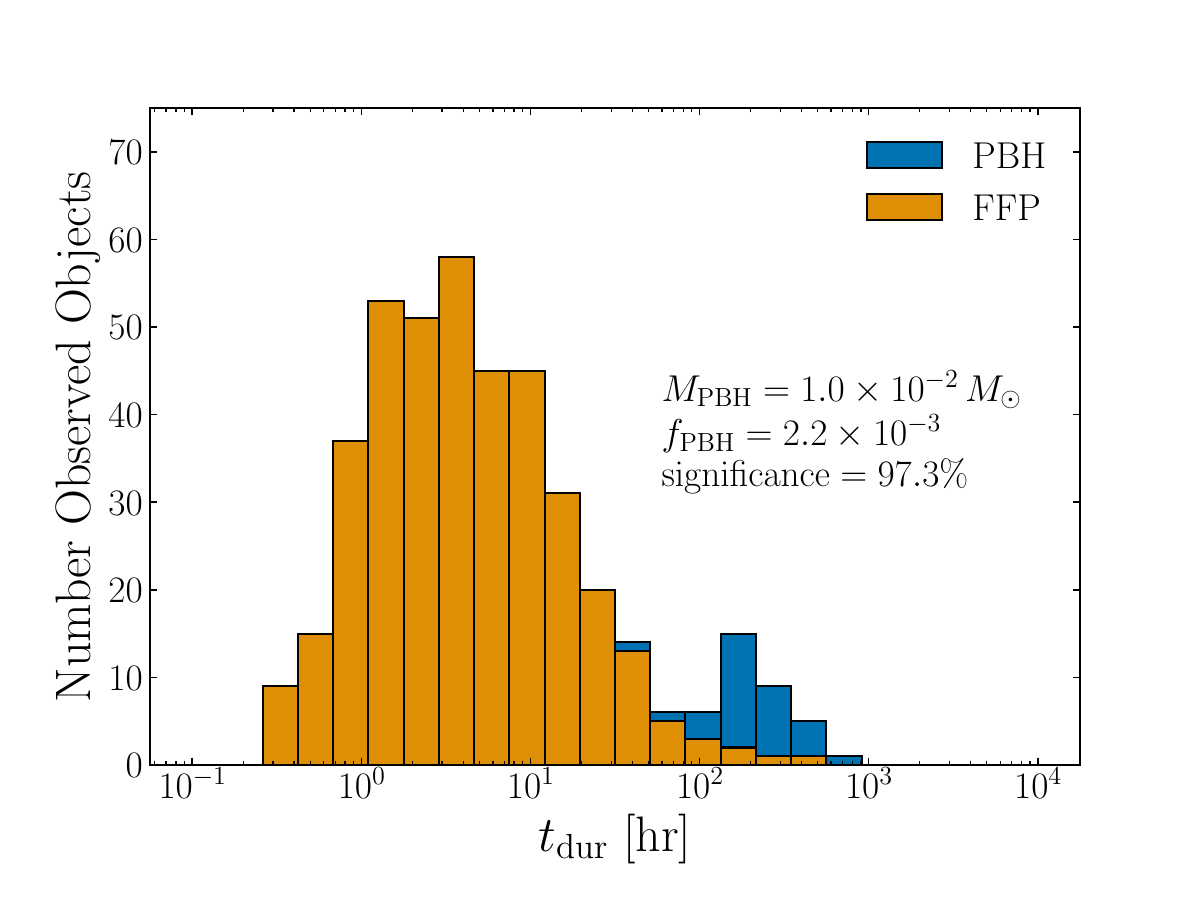}
\caption{ A stacked histogram of FFP and FFP+PBH distributions that are \textit{distinguishable} at 95\% confidence. These distributions correspond to parameter values of $\mathcal{N} = 10$, $p = 1.0$.  The associated observable yields at this point in parameter space are  $N_{\rm{FFP}} = 389$, $~N_{\rm{PBH}} = 8\% $ $N_{\rm{FFP}}$.}
\label{fig:ad1d_succeed}
\end{minipage}\hfill
\begin{minipage}[t]{0.48\textwidth}
\centering
\vspace{0pt}
\includegraphics[width=\textwidth]{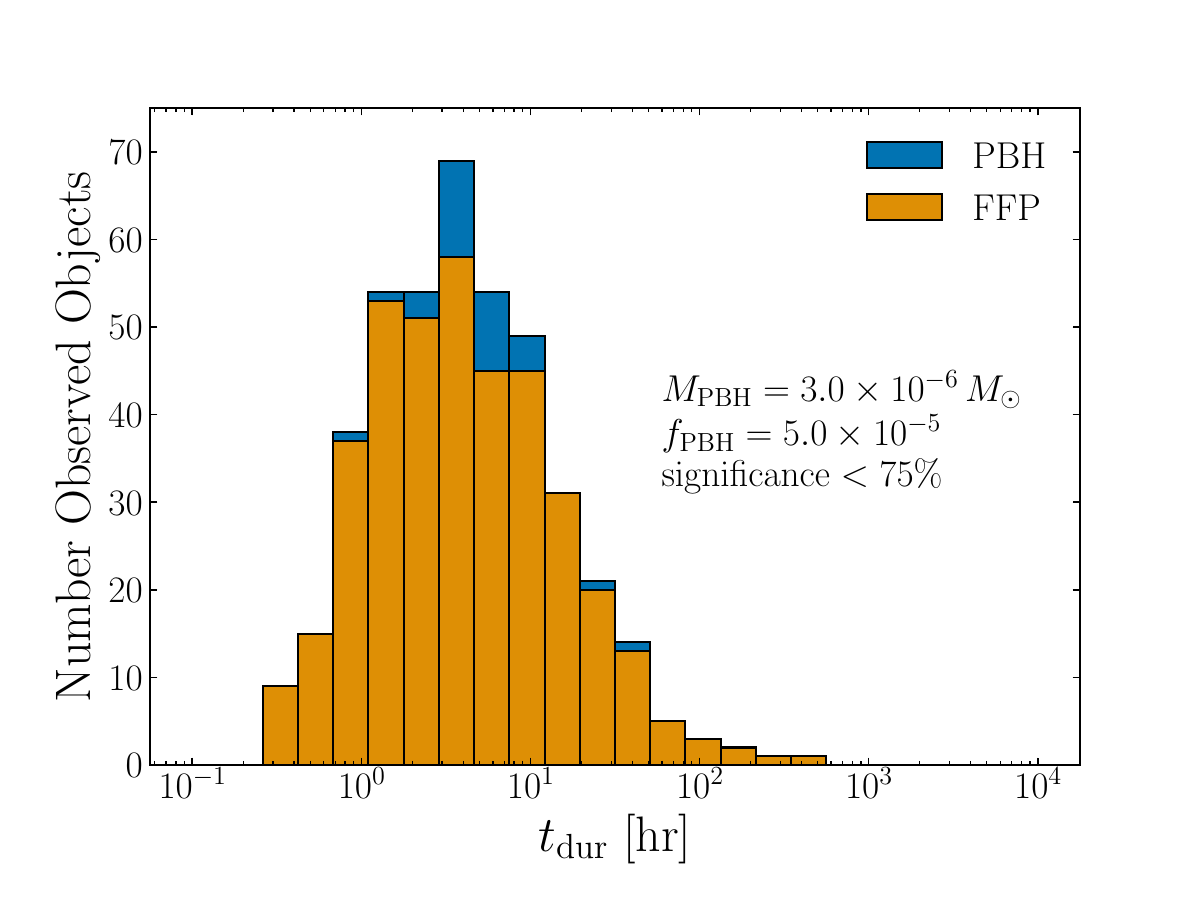}
\caption{ A stacked histogram of FFP and FFP+PBH distributions that are \textit{indistinguishable} at 95\% confidence. These distributions correspond to parameter values of $\mathcal{N} = 10$, $p = 1.0$. These parameters were chosen to yield the same observable yields as Fig. \ref{fig:ad1d_succeed}, $N_{\rm{FFP}} = 389$, $~N_{\rm{PBH}} = 8\%$ $N_{\rm{FFP}}$, however with a different location of the PBH peak.}
\label{fig:ad1d_fail}
\end{minipage}
\end{figure*}

\subsection{Subpopulation identification}
\label{subsec:AD}
Our statistical analysis relies upon discriminating between $t_{\rm{dur}}$ distributions sourced by either purely FFPs or a combination of FFPs and PBHs. We will define the true distributions from which a particular set of detected events are sampled as $\mathcal{T}_\text{dur}^\text{FFP}$ and $\mathcal{T}_\text{dur}^\text{FFP+PBH}$.
These distributions depend on a complex combination of several input parameters, including the power-law index of FFPs ($p$), the central mass of the PBH distribution ($M_{\rm{PBH}}$), and the overall number of observed FFPs and PBHs ($N_{\rm{FFP}}$ and $N_{\rm{PBH}}$). As such, they cannot be computed in a closed analytic form. We therefore choose to employ a test that discriminates based purely on empirical distribution functions without relying on an underlying analytic background model.
The two-sample Anderson-Darling (AD) test is an effective choice for this situation\footnote{In practice, Roman will likely perform a Bayesian analysis to estimate the parameters controlling the lens distribution, which will be more sensitive than the methodology we employ here. However, the AD test provides a robust, if conservative, estimate of Roman's sensitivity.}, as it is non-parameteric, hence requires no model input, and outperforms the Komolgorov-Smirnov test in the amount of data required for significance, see \cite{ADKS}.

The AD test computes the significance at which two test distributions are sampled from the same underlying distribution. Given two distributions of size $m$, $n$ sampled from the true distributions $\mathcal{T}_\text{dur}^\text{FFP}$ and $\mathcal{T}_\text{dur}^\text{FFP+PBH}$, we construct two empirical distribution functions, denoted $\mathcal{T}_{\text{dur},m}^\text{FFP}$ and $\mathcal{T}_{\text{dur},n}^\text{FFP+PBH}$, respectively. In the context of our analysis, $m = N_\text{FFP}$ and $n = N_\text{FFP} + N_\text{PBH}$, where $N_\text{FFP}$ and $N_\text{PBH}$ are calculated as described in the previous subsection. In terms of these empirical distribution functions, the AD test statistic can be written as \cite{scholz-ad}:
\begin{equation}
\label{eq:ad_statistic}
    A_{m n}^2 = \frac{m n}{N} \int^{\infty}_{-\infty} \frac{(\mathcal{T}_{\text{dur}, m}^\text{FFP} -\mathcal{T}_{\text{dur}, n}^\text{FFP+PBH})^2}{\mathcal{K}_N (1 - \mathcal{K}_N)} \, d\mathcal{K}_N
\end{equation}
where 
\begin{equation}
    \mathcal{K}_N = \frac{1}{N} (m \mathcal{T}_{\text{dur}, m}^\text{FFP} + n \mathcal{T}_{\text{dur}, n}^\text{FFP+PBH})
\end{equation}
and $N \equiv m + n$. Note that by performing this test, we do not necessarily learn the PBH mass or abundance; merely that the distributions are separable.

To determine the sensitivity, we fix $\mathcal{N}$, $p$, $M_\text{PBH}$, and $\sigma$ and allow $r \equiv N_\text{PBH}/N_\text{FFP}$ to vary. We set our limit at the value of $r$ such that the AD test rejects the null hypothesis (i.e. both distributions are sampled from a pure FFP distribution) 
at 95\% confidence. Representative examples of distributions that are distinguishable and indistinguishable by the AD test are displayed in Figs. \ref{fig:ad1d_succeed} and \ref{fig:ad1d_fail}, respectively. In Fig. \ref{fig:ad1d_succeed}, the PBH distribution peaks at $t_\text{dur}$ values well above the majority of FFPs, hence is readily distinguishable. In Fig. \ref{fig:ad1d_fail}, despite having the same number of observed FFPs and PBHs as in Fig. \ref{fig:ad1d_succeed}, the two peaks overlap and the PBH population cannot be discriminated from background.

The weakness of this test is that in the low-statistics regime, two distributions may appear to have been drawn from different underlying distributions purely due to random fluctuations. In order to mitigate this effect, we perform our analysis 10 times and take the mean of the results, which we have verified numerically is sufficient for suppressing statistical fluctuation throughout our parameter space.

The analysis described above solely sets a limit on $r$, the \textit{ratio} of observed PBH yield to FFP yield. In order to connect this to a physical density, we must calculate these yields. To do so, we employ \texttt{LensCalcPy} and produce two reference yield curves. The first is the expected yield of observable PBHs as a function of $M_\text{PBH}$ for $f_\text{PBH} = 1$, which we denote $N_\text{PBH}^{f=1} ( M_\text{PBH})$ and appears in Fig. \ref{fig:pbh_yield}. The second is the expected yield of observable FFPs for $\mathcal{N} = 10$ as a function of $p$, which we denote $N_\text{FFP}^{\mathcal{N} = 10}(p)$ and appears in Fig. \ref{fig:ffp_yield}.  The $f_\text{PBH}$ corresponding to a particular $r$ is therefore simply given by $f_\text{PBH}(M_\text{PBH},p) = r \times [N_\text{FFP}^{\mathcal{N}=10}(p)/N_\text{PBH}^{f=1} ( M_\text{PBH})]$.

Our results depend implicitly on $\mathcal{N}$ and $p$, the true values of which are unknown. Existing observations suggest possible values in the range $p \approx 0.66 - 1.33$ and $\mathcal{N} \approx 2 - 20$ \cite{sumi2023freefloating,gould_free-floating_2022,mroz2023exoplanet,Johnson_2020}. We therefore choose to marginalize over this uncertainty by determining, for a given $M_\text{PBH}$ and $\sigma$, the $p \in [0.66, 1.33]$ for which our analysis is weakest and adopting the corresponding $f_\text{PBH}$ as our limit. To capture the uncertainty on $\mathcal{N}$, we choose to display three results: our fiducial results ($\mathcal{N}$ = 10), as well as results in which $\mathcal{N}$ has been taken to be larger/smaller than our fiducial value by an order of magnitude. This likely dramatically overestimates the uncertainty on this parameter given current constraints. However, by adopting this range, we encapsulate both the intrinsic uncertainty on its value as well as the uncertainty induced by our Galactic model (see App. \ref{sec:yields}).

\section{Results and Discussion}
\label{sec:disc}

\begin{figure}[b]
\includegraphics[width=\linewidth]{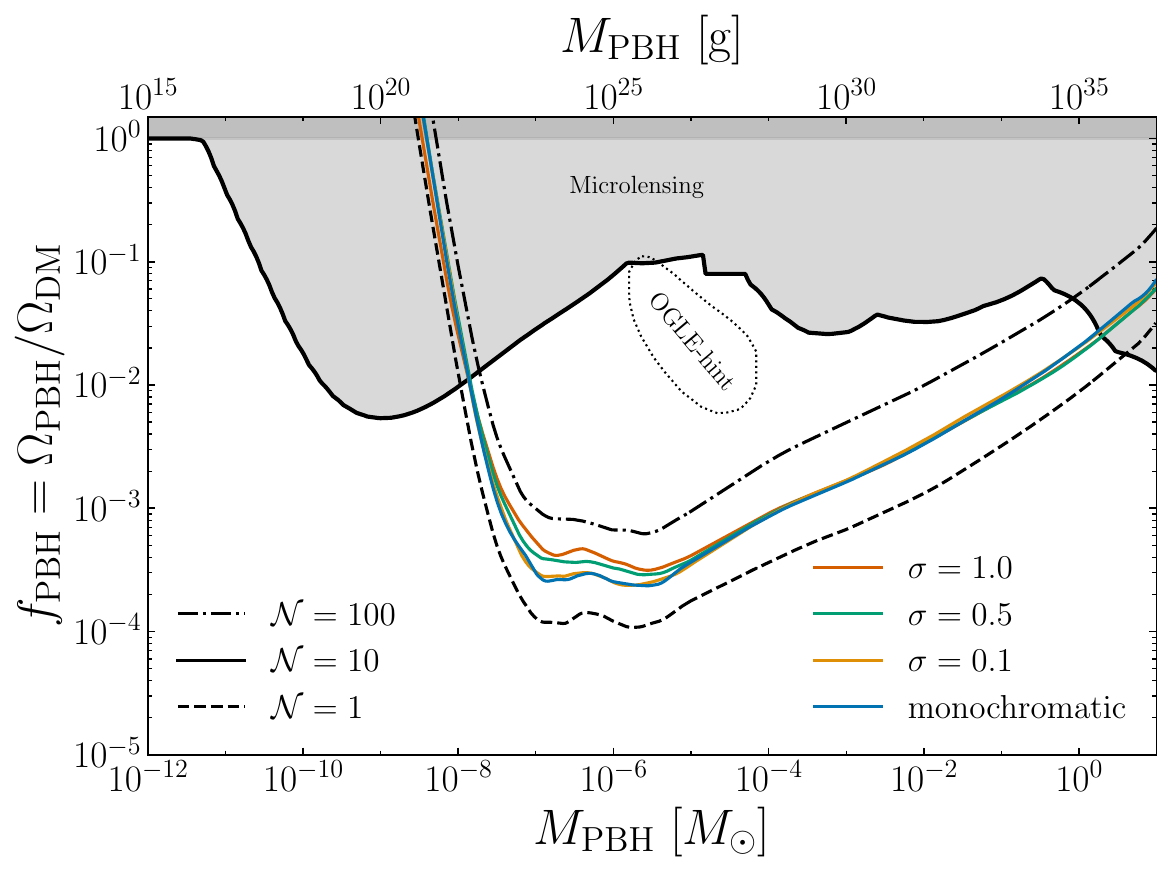}
\caption{\label{fig:s} Roman sensitivity to detecting a population of PBHs in a background of FFPs. The solid curves correspond to $\mathcal{N} = 10$ and varying width $\sigma$ of the log-normal PBH distribution, while the dashed and dot-dashed curves correspond to $\mathcal{N} = 1$ and $100$, respectively. Existing constraints on the PBH abundance are shown in gray \cite{bradley_j_kavanagh_2019_3538999} and the region in which existing observations hint at a population of PBHs \cite{Niikura_2019} is denoted ``OGLE hint.'' See text for details.}
\end{figure}

We display our ultimate sensitivity curves in Fig.~\ref{fig:s}. Existing constraints are shown in gray \cite{bradley_j_kavanagh_2019_3538999}. Additionally, we have included a dotted region (``OGLE hint'') corresponding to the parameter space in which the short-timescale events observed by OGLE can be explained by a population of PBHs \cite{Niikura_2019}. The solid curves correspond to a fiducial FFP normalization of $\mathcal{N} = 10$ and varying width of the log-normal PBH distribution, while the dashed and dot-dashed curves correspond to $\mathcal{N} = 1$ and 100, respectively for a monochromatic PBH mass distribution. As described in Sec. \ref{subsec:AD}, these extreme values of $\mathcal{N}$ likely significantly overestimate the uncertainty on the FFP distribution, however, as can be seen in Fig. \ref{fig:s}, even these variations only induce changes to the sensitivity at the sub-magnitude level. Note that the largest number density of FFPs corresponds to the \textit{weakest} sensitivity, as a larger FFP yield requires a correspondingly larger PBH yield to achieve the same significance of discrimination. All curves displayed have been marginalized over $p$ via the methodology described in \ref{subsec:AD}.

Roman's sensitivity to identifying a subpopulation of PBHs peaks at $f_\text{PBH}\sim 10^{-4}$ in the mass range $M_\text{PBH} \approx 10^{-8}\,M_\odot - 10^{-6}\,M_\odot$. Both the location of this peak and the corresponding value of $f_\text{PBH}$ can be understood simply. Since the number density of PBHs scales as $1/M_\text{PBH}$ for fixed $f_\text{PBH}$, the location of peak sensitivity corresponds to the lowest possible mass before finite-source effects reduce detectability. As discussed in Sec. \ref{sec:microlensing}, finite-source effects become relevant when $\theta_S \approx \theta_E$, a condition that can be rewritten in terms of mass to yield \cite{derocco_constraints_2023}
\begin{equation}
    M_\text{finite} \approx \frac{\theta_S^2 c^2 d_L}{4 G (1- \frac{d_L}{d_S})}\Big(\frac{d_L}{d_S}\Big).
\end{equation}
Assuming the source to have a radius comparable to that of the Sun and taking $d_S = 8.5 ~\rm{kpc}$ and $d_L = 7.0 ~\rm{kpc}$ as typical distances for lensing events in the Galactic Bulge, one finds $M_\text{finite} \approx 10^{-6} M_{\odot}$, which corresponds with the mass at which the sensitivity peaks in Fig. \ref{fig:s}.

Similarly, $f_\text{PBH}$ can be estimated at this peak. We find that at terrestrial masses, a PBH yield of roughly 10\% $N_\text{FFP}$ is sufficient to identify the PBH subpopulation. Figs.~\ref{fig:pbh_yield} and \ref{fig:ffp_yield} show that Roman's expected yield for FFPs and PBHs (at $\mathcal{N} = 10$ and $f_\text{PBH} = 1$) are $\mathcal{O}(1000)$ and $\mathcal{O}(10^6)$, respectively. We therefore see immediately that $N_\text{PBH} \approx 10\% \, N_\text{FFP}$ corresponds to $f_\text{PBH} \sim 10^{-4}$, which matches onto the maximal sensitivity shown in Fig. \ref{fig:s}.

In the region of peak sensitivity, we find that sensitivity weakens with increasing width of the log-normal PBH distribution. This is not due to the fact that broader PBH distributions appear more akin to the FFP power law, but rather because broadening the PBH distribution pushes PBHs outside the observable window and lowers the overall yield of observable PBH events. This can be seen in Fig. \ref{fig:pbh_yield}, where broadening the distribution causes a monotonic decrease in the number of detected events in the region of peak sensitivity.\footnote{Note that well outside this region, the opposite effect can actually improve sensitivity marginally for broad distributions by pushing events into the observable window.}
For a fixed number of PBHs required for discrimination, this reduced detection rate must be compensated for by an increase in $f_{\text{PBH}}$.

The small decrease in sensitivity at $M_\text{PBH}\approx 10^{-7}\, M_\odot$ is due to the peak of the PBH $t_\text{dur}$ distribution coinciding with the peak of the FFP $t_\text{dur}$ distribution, as can be seen in Fig.~\ref{fig:ad1d_fail}. At slightly higher and lower $M_\text{PBH}$, the two distributions peak at slightly different $t_\text{dur}$, improving sensitivity. However, this effect is small, as the sensitivity is predominantly governed by the PBH yield, which decreases rapidly at masses much above $10^{-6}\, M_\odot$ and below $10^{-8}\,M_\odot$.

In summary, our results show that even under conservative assumptions about Roman's detection threshold (Sec. \ref{subsec:eventrate}) and the underlying background of FFPs (Sec \ref{subsec:AD}), the Galactic Bulge Time Domain Survey will be highly sensitive to detecting a population of PBHs in new regions of parameter space. Excitingly, these regions include the parameter space in which existing short-timescale events have been suggested to hint at a subpopulation of PBHs at terrestrial masses \cite{Niikura_2019}. Roman is therefore poised not only to make the first precise measurements of the FFP mass distribution, but to possibly uncover a subpopulation of PBHs lying within it as well.

\section{Conclusions}
\label{sec:conc}

The launch of the Nancy Grace Roman Space Telescope will open a new window into low-mass astrophysical bodies. Though its Galactic Bulge Time Domain Survey targets bound and unbound exoplanets, we have shown that it will have unprecedented sensitivity to physics beyond the Standard Model as well. In particular, it will probe the fraction of dark matter composed of primordial black holes at abundances as low as $f_\text{PBH} \approx 10^{-4}$ at PBH masses of roughly $10^{-6} \, M_\odot$, with a sensitivity that decreases as $\approx M_\text{PBH}^{1/3}$ towards higher masses. Its region of sensitivity extends up to three orders of magnitude below existing constraints. This region fully encompasses the parameter space in which an excess of short-duration microlensing events observed by OGLE have been suggested to hint at a population of PBHs \cite{Niikura_2019}. Therefore, Roman will conclusively determine the nature of these events, whether it be rogue worlds or our first glimpse of what lies on the dark side of the universe.

% -----------------------------------------------------------------------------
\vspace{0.2cm}
\noindent {\it Acknowledgements.}
%-----------------------------------------------------------------------------
The authors wish to thank Scott Gaudi and Samson Johnson for useful input on Roman's sensitivity and clarification on estimates of Roman's yield. They would also like to thank Duncan Wood for providing insight on microlensing opportunities in other upcoming surveys. WD would like to thank Masahiro Takada for bringing up future opportunties to probe the PBH population with alternate lines of sight. This work was supported in part by DOE grant No. DE-SC0010107.
% \bibliography{references.bib}
% \pagebreak

\appendix

\section{Comparison of Estimated Yields}
\label{sec:yields}

In this Appendix, we compare the fiducial FFP yield calculated in our analysis 
to that of \cite{Johnson_2020}.
The authors of \cite{Johnson_2020} calculate their expected FFP yield for the Roman GBTDS using the code \textit{Gravitational microlensing Using Large Lensed Sources} (GULLS) \cite{penny_predictions_2019}. GULLS draws explicit sources and lenses from a Bescançon galactic model (version 1106 \cite{10.1093/mnrasl/slad049}) and simulates individual microlensing events by generating realistic photometry using synthetic images. This approach is significantly different from the semi-analytic approach we employ in our paper. \texttt{LensCalcPy}, the code used to compute our FFP yields, is designed to provide simple estimates of lensing event rates, not to model individual events or generate associated photometry. 
However, its speed and flexibility makes it well-suited to population-level studies with large numbers of events. 

While our approach and that of \cite{Johnson_2020} differ significantly in implementation, we find that they produce very similar ultimate FFP yields. In order to see this, we compare to Table 2 of \cite{Johnson_2020}, where the authors have displayed their fiducial FFP yield for a log-uniform mass distribution ($\frac{dN}{d \log{M}} = 1 ~\rm{dex}^{-1}$) as a function of FFP mass. Performing the equivalent analysis with \texttt{LensCalcPy} and adopting the normalization of $1~\text{dex}^{-1}$ results in the yields shown in Table \ref{tab:johnson_compare}. We see that at masses $> M_\oplus$, our yields differ from those of \cite{Johnson_2020} by less than a factor of two. At lower masses, the discrepancy between the approaches grows, reaching a value of $\approx 6$ at the lowest observable masses.

\begin{table}[htbp]
    \centering
    \caption{FFP yield comparison for Log-Uniform Mass distribution}
    \label{tab:johnson_compare}
    % Enclose the table in a box
    \begin{tabular}{|c|c|c|}
        \hline
        \textbf{Mass} $(M_{\oplus})$ & {\textbf{Johnson et al.} \cite{Johnson_2020}} & {\textbf{This work}} \\
        \hline
        0.01   & 0.31  & 0.05     \\
        \hline
        0.1    & 4.49  & 1.75     \\
        \hline
        1      & 22.1  & 19.0      \\
        \hline
        10     & 87.1   & 72.6      \\
        \hline
        100    & 313  & 234     \\
        \hline
        1000   & 1025  & 744     \\
        \hline
        10,000 & 3300 & 2370    \\
        \hline
    \end{tabular}
\end{table}

\begin{figure}[b]
\includegraphics[width=\linewidth]{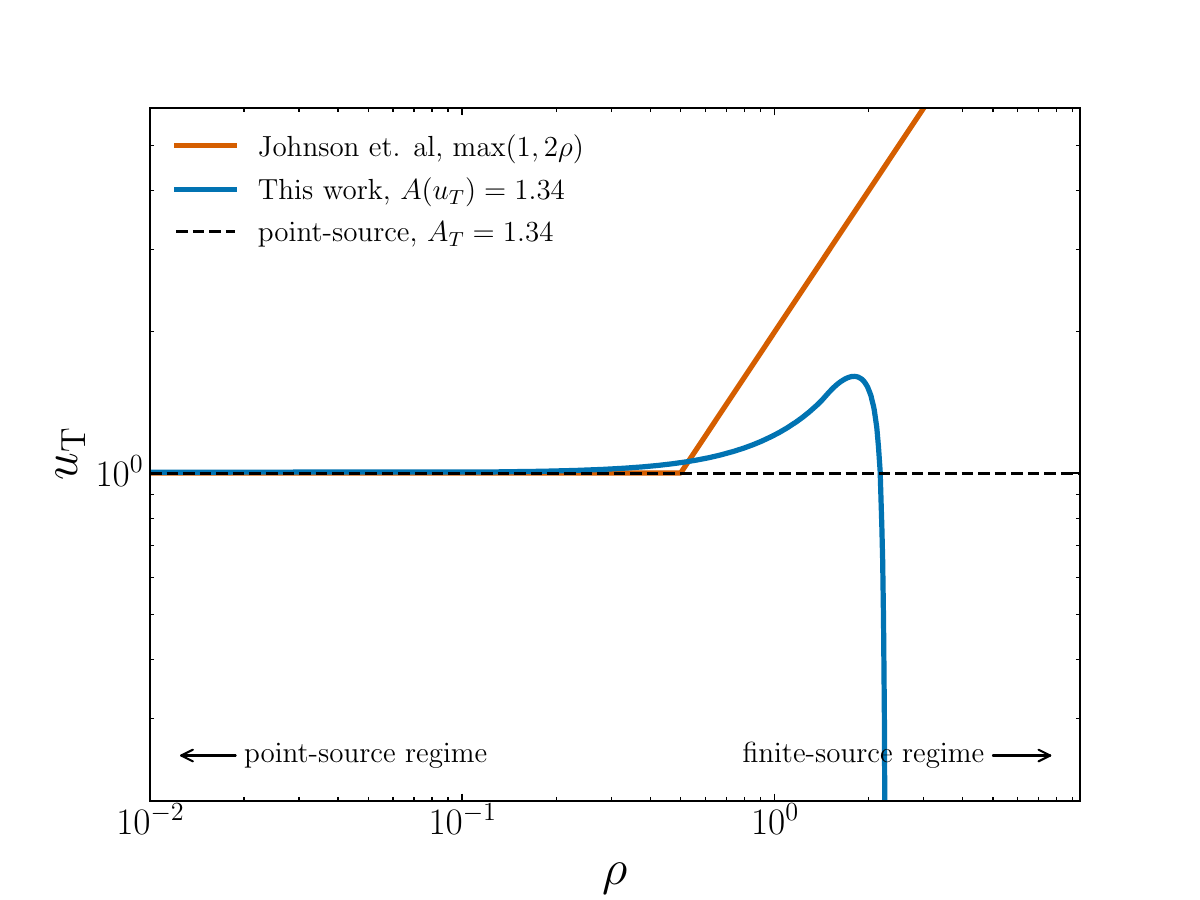}
\caption{The threshold impact parameter as a function of $\rho = \theta_S/\theta_E$. The methodology of Johnson et al. \cite{Johnson_2020} (orange) results in larger threshold impact parameters in the finite-source regime than our analysis (blue), increasing their relative yields.}
\label{fig:ut_compare} 
\end{figure}

We see that our results tend to underestimate the total FFP yield compared to GULLS, particularly for low-mass objects. A primary source of this discrepancy stems from differences between the definition of maximum detectable impact parameter in the two analyses, which we compare in Fig.~\ref{fig:ut_compare}. In \cite{Johnson_2020}, $u_{\rm{min}}$ is drawn uniformly from [$0, \rm{max}(1, 2 \rho)$] when generating an event. This effectively sets 
\begin{equation}
    u_T = 
    \begin{cases}
        1 & \rho < 0.5~\text{(point-source regime)}\\
        2\rho & \rho > 0.5~\text{(finite-source regime)}
    \end{cases}
\end{equation}
resulting in the orange curve shown in Fig. \ref{fig:ut_compare}. As described in Sec. \ref{subsec:eventrate}, in our analysis, we instead determine the maximal impact parameter by solving the implicit equation $A_\text{finite}(u_T, \rho) = A_T$.  This yields the blue curve in Fig. \ref{fig:ut_compare}. We choose to adopt $A_T = 1.34$ as our fiducial threshold throughout our analysis. This agrees with \cite{Johnson_2020} in the point-source regime, however in the finite-source regime (which is most relevant for low-mass objects), their approach yields generically larger values of $u_T$ than ours, as can be seen in Fig. \ref{fig:ut_compare}. Thus, their effective threshold magnification is $< 1.34$, resulting in the increased yields at low masses seen in Table \ref{tab:johnson_compare}. While we have chosen to use $A_T = 1.34$ throughout our analysis, this is likely an underestimate of Roman's ultimate detection threshold, which has been suggested to reach values of $\lesssim 1\%$ increases in flux for sufficiently bright sources \cite{Johnson_2020}.
We therefore note that depending on the photometric sensitivity achieved by Roman, our current yield predictions may underestimate the number of detected FFP events. This uncertainty is, however, encapsulated by the range of normalizations in the mass functions considered and thus in the curves shown in Figure \ref{fig:s}.

\pagebreak
\normalem
\bibliography{references.bib}

\onecolumngrid

\end{document}